\documentclass[aps,prl,twocolumn,superscriptaddress,showpacs]{revtex4}

\usepackage{graphicx}
\usepackage{afterpage}

\begin{document}


\title{Fermi surface of Cr$_{1-x}$V$_x$ across the quantum critical point}



\author{J.F. DiTusa}
\email[]{ditusa@phys.lsu.edu}
\affiliation{Department of Physics and Astronomy, Louisiana State
University, Baton Rouge, Louisiana 70803, USA}

\author{R. G. Goodrich}
\affiliation{Department of Physics, George Washington
University, Washington, DC 20052, USA}

\author{N. Harrison}
\affiliation{National High Magnetic Field Laboratory, Los Alamos
National Laboratory, MS E536, Los Alamos, New Mexico 87545, USA}

\author{E. S. Choi}
\affiliation{National High Magnetic Field Laboratory, Florida State
University, Tallahassee, Florida 32310, USA}


\date{\today}

\begin{abstract}
We have measured de Haas-van Alphen oscillations of Cr$_{1-x}$V$_x$,
$0 \le x \le 0.05$, at high fields for samples on both sides of the
quantum critical point at $x_c=0.035$. For all samples we observe only
those oscillations associated with a single small hole band with
magnetic breakdown orbits of the reconstructed Fermi surface evident
for $x<x_c$. The absence of oscillations from Fermi surface sheets
most responsible for the spin density wave (SDW) in Cr for $x>x_c$ is
further evidence for strong fluctuation scattering of these charge
carriers well into the paramagnetic regime. We find no significant
mass enhancement of the carriers in the single observed band at any
$x$. An anomalous field dependence of the dHvA signal for our
$x=0.035$ crystal at particular orientations of the magnetic field is
identified as due to magnetic breakdown that we speculate results from
a field induced SDW transition at high fields.
\end{abstract}


\maketitle

\section{Introduction}
Chromium has fascinated condensed matter physicists since it was
demonstrated to be antiferromagnetic, AFM, by Shull and
Wilkinson\cite{shull} in 1953.  The understanding of the spin density
wave, SDW, ground state and its ramifications on the physical
properties of this elemental metal was largely worked out over 30
years ago\cite{fawcettrev,fawcettrev2}. However, more recently it was
realized that the continuous suppression of the N\'eel temperature,
$T_N$, to zero by the substitution of small amounts of V, Nb, or Ta,
or by application of pressure made Cr an ideal simple material to
explore the consequences of a quantum criticality\cite{yeh,lee}. The
results of investigations into this transition have been used to
compare to more complicated systems where quantum critical points
(QCP) are thought to dominate the physical properties over wide ranges
of composition, temperature, pressure, and magnetic field. The
association of superconductivity with QCPs has also fueled the
interest in exploring such transitions. One well established method
for accessing the Fermi surfaces, FS, of metals is the de Haas van
Alphen, dHvA, effect\cite{shoenberg} which is increasingly being
employed to explore FS changes in materials near
QCPs\cite{capan,goh,jaudet,settai,goodrich2,rourke}. Here we measure
the dHvA oscillations in Cr$_{1-x}$V$_x$ as $x$ is varied through the
critical concentration at
$x=0.035$\cite{fawcettrev,fawcettrev2,yeh,lee}. Our data near the
simplest of QCPs should serve as a point of reference for the more
complex quantum critical systems and may have consequences for their
interpretation. Our experiments make use of very large magnetic fields
that were not available to the earlier
explorations\cite{graebner,fawcett,venema,ruesink} to extend the dHvA
measurements through $x_c$ where the doping induced disorder makes
small cyclotron orbits a requirement for observation.

The incommensurate SDW state in Cr differs from classical AFMs as
magnetic moments at the cube corners of the body centered cubic (bcc)
structure are antiparallel, but not quite equal to the magnetic
moments on the body-center. Their magnitude is modulated in a
sinusoidal manner with a wavevector $Q_{SDW}$ which lies along a [100]
crystal axis. In pure Cr $Q_{SDW}=0.953 a^*$ at liquid helium
temperature and is accompanied by a charge density wave modulation at
$2Q_{SDW}$\cite{tsunoda}.  Band structure calculations of the Fermi
surface of paramagnetic (PM) bcc Cr is very similar to that of Mo and
W, the other elements of column VIB of the periodic
table\cite{lomer,fry,mattheiss}.  These FSs are typically described as
consisting of an electron jack centered at $\Gamma$, a hole octahedron
centered on the H-points [$\pi/a$,0,0], hole ellipsoids centered at
the N-points [$\pi/2a$,$\pi/2a$,0], and small electron lenses along
the $\Gamma$-H line of the conventionally labeled Brillouin zone. In
Cr extensive nesting between the electron jack and the hole octahedron
is responsible for the transition to the SDW state as $Q_{SDW}$
connects these two parts of the FS so that they are gapped below $T_N
= 313$ K.  As a result only smaller sheets of the Fermi surface are
expected to survive in the SDW phase in agreement with Hall effect
experiments that reveal a substantial decrease in itinerant carriers
below $T_N$\cite{yeh,lee,devreis}. V substitution reduces the electron
density so that electron-like regions of the Fermi surface are
expected to shrink in size while the hole-like regions expand. The
result is a decreased area on the electron jack and hole octahedron
which remain parallel and a decreased $Q_{SDW}$ corresponding to an
increased incommensurability with $x$. The area of parallel Fermi
surfaces connected by $Q_{SDW}$ appears to drop quickly for
$x>0.03$\cite{trego} and the QCP is reached when the interaction area
between the two FS sheets is reduced to the point where AFM cannot be
sustained.

The periodicity caused by the condensation into the SDW phase also has
dramatic consequences for the other FS sheets as the PM FS is remapped
by translation through $\pm nQ_{SDW}$, where $n$ is an
integer\cite{lomer,fry}. For the pockets of holes at the N-points this
results in extensive overlapping hole ellipsoids (demonstrated
schematically in Fig.~\ref{fig:tfft} below) and, thus, open orbits
along the direction of $Q_{SDW}$ that have been detected in
magnetoresistance and Hall effect measurements in single domain
samples\cite{arko,reifenberger}. Previous dHvA measurements in Cr
revealed a large number of small $k$-space area orbits caused by the
magnetic breakdown between these intersecting hole surfaces. No other
pieces of the Fermi surface have been successfully observed with this
method in contrast to dHvA experiments performed on Mo\cite{hoekstra}
and W\cite{girvan} where the larger FS sheets are
apparent. Angle-resolved photoemission (ARPES) measurements of the
electronic structure of Cr$_{1-x}$V$_x$ thin
films\cite{krupin,schafer} observe all of the dispersing bands and
indicate a complete gapping of the electron jack and hole octahedron
in the SDW phase.  Measurements of the FS of Cr$_{1-x}$V$_x$ for $0\le
x\le 1$ via two dimensional angular correlation of electron-positron
annihilation radiation techniques (2D-ACAR) observed each of the FS
sheets and carefully followed the evolution of the pocket of holes at
the N-points for comparison to measurements of the giant
magnetoresistance effect in multilayers with Cr
spacers\cite{hughes}. dHvA measurements in dilute V substituted
samples ($x<0.01$) observed small changes consistent with a reduction
of $Q_{SDW}$ with $x$ without a measurable change to the cross
sectional area of the N-pockets of the Fermi surface\cite{gutman}. In
addition dHvA measurements on Mn doped Cr, where the SDW phase becomes
commensurate for concentrations of 1\% or larger\cite{koehler}, reveal
a smaller number of oscillation frequencies corresponding to a much
simpler Fermi surface.

Here we find that V substitution leads to changes in the observed
frequencies in the dHvA spectra consistent with an increased area of
the hole ellipsoid at the N-point along with a decreased $Q_{SDW}$ for
$x<x_c$. For $x\ge x_c$ we observe a much simpler dHvA spectra as only
signals associated with a {\it single} ellipsoid are apparent. We
interpret the absence of observed frequencies associated with the
larger FS sheets for PM Cr as indicating that the scattering of these
carriers due to the SDW fluctuations remains strong, even for
$x=0.05$. This is consistent with previous transport
measurements\cite{yeh,lee} showing little change in the low
temperature carrier conductivity. In addition, for $x=0.035$ we
observe changes to the spectra that indicate the existence of magnetic
breakdown orbits for fields larger than $\sim 20$ T. We speculate that
these orbits may be indicating a return to the SDW state due to the
changes in the FS geometry with field, a field induced SDW transition
for samples just on the PM side of the QCP.

\section{Experimental Details}
Single crystals of Cr, Cr$_{0.98}$V$_{0.02}$, Cr$_{0.965}$V$_{0.035}$,
and Cr$_{0.95}$V$_{0.05}$ were grown at Ames laboratory from high
purity starting materials by arc melting, arc zone refinement, and
heat treatment at 1600$^o$C for 72 hours. These crystals were oriented
by back reflection Laue and were spark cut so that faces were
perpendicular to [100] directions. The resulting single crystals had
approximate dimensions of 0.25 mm $\times$ 0.25 mm $\times$ 0.1 mm
thick. Resistivity measurements of our nominally pure and $x=0.02$
crystals displayed anomalies associated with $T_N$ within 1 K of those
found in previous measurements\cite{arajs} of samples with the same
stoichiometry. No anomaly was found for our $x=0.035$ and 0.05 samples
indicating, as expected, that $x>x_c$. Some of the dHvA measurements
were made between 0 and 55 T at the pulsed field facility of the
National High Magnetic Field Laboratory (NHMFL) located at the Los
Alamos National Laboratory. Angular dependent measurements were made
between 0 and 33 T at the NHMFL branch located in Tallahassee, FL
using a cantilever to measure the torque on the sample as the field
was changed. Details of both of these measurement techniques have
previously been reported\cite{goodrich1}.

\section{de Haas-van Alphen Spectra}
\begin{figure}[htb]
  \includegraphics[angle=90,width=3.0in,bb=50 280 550
  677,clip]{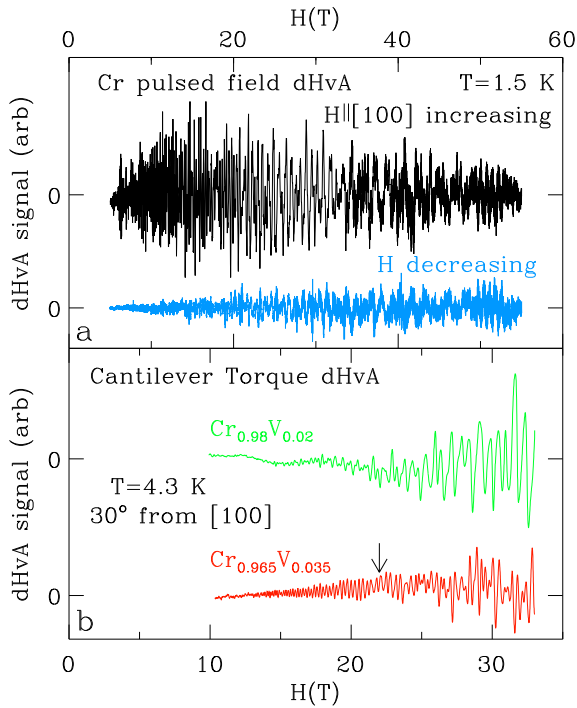}%
  \caption{\label{fig:rawdata} (color online) de Haas-van Alphen
oscillations from the pulsed field and torque cantilever measurements.
a) Pickup coil signal for the pulsed field measurement of a nominally
pure Cr crystal oriented with the magnetic field, $H$, parallel to the
crystal [100] direction. The data were taken at 1.5 K and the signal
is shown for both increasing and decreasing magnetic fields.  b)
Torque signal from the cantilever measurements of the dHvA
oscillations for a Cr$_{0.98}$V$_{0.02}$ and a Cr$_{0.965}$V$_{0.035}$
single crystal.  Data taken at 4.3 K with the magnetic field oriented
30$^o$ from the [100] crystal axis. Arrow indicates the field (22 T)
where magnetic breakdown is evident in the $x=0.035$ sample at this
orientation, see text.}
\end{figure}

\begin{figure}[htb]
  \includegraphics[angle=90,width=3.5in,bb=65 75 534
  677,clip]{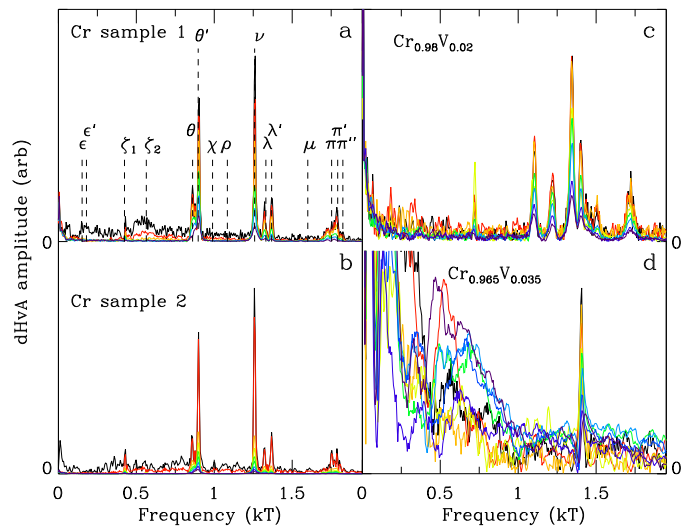}%
  \caption{\label{fig:pfft} Fourier transforms of the de Haas-van
Alphen oscillations from the pulsed field measurements. a) and b)
Fourier transforms (FFT) of de Haas-van Alphen, dHvA, signal for two
Cr crystals at temperatures, $T$, of 1.45 (black), 2.5 (red), 4.2
(orange), 8.0 (green), 12.0 (blue), and 16 K (violet) for (a) and 0.45
(black), 1.45 (red), 4.2 (orange), 6.0 (yellow), 8.0 (yellow-green),
10.0 (green), 13.0 (blue), 16.0 (purple), and 20 K (violet) in (b). c)
FFT of dHvA signal for Cr$_{0.98}$V$_{0.02}$ at $T$s of 0.85 (black),
1.5 (red), 2.4 (orange), 4.2 (yellow), 6.0 (green), 8.0 (blue), 12.0
(purple), and 14.0 K (violet). d) FFT of the dHvA signal for
Cr$_{0.965}$V$_{0.035}$ at $T$s of 0.50 (black), 1.5 (red), 2.0
(orange), 4.2 (yellow), 6.0 (green), 8.0 (blue), 10.0 (dark blue),
12.0 (purple), and 14.0 K (violet). All data taken with the magnetic
field, $H$, parallel to a [100] crystal axis with FFTs calculated for
fields between 15 and 55 T. The dashed lines in (a) indicate the dHvA
frequencies reported in Ref.~{\protect {\cite{venema,ruesink}}} and we
employ the labels adopted previously for these frequencies.  }
\end{figure}

\begin{figure}[htb]
  \includegraphics[angle=90,width=3.5in,bb=0 0 534
  677,clip]{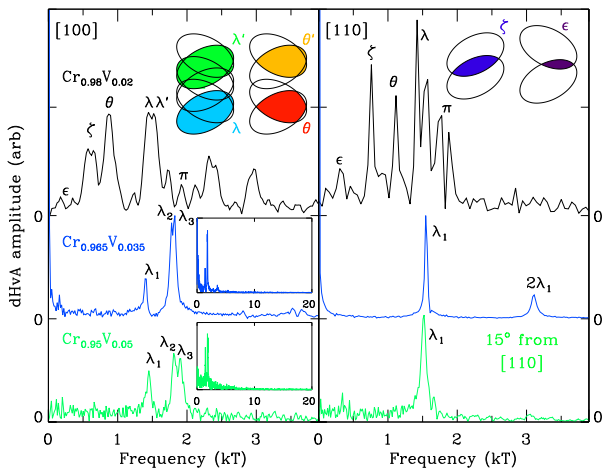}%
  \caption{\label{fig:tfft} Fourier transforms of the de Haas-van
Alphen oscillations from torque cantilever measurements. FFT of the
dHvA signal for fields parallel to a [100] crystal axis (left side)
and parallel to a [110] crystal axis (right side) for
Cr$_{0.98}$V$_{0.02}$ (black), Cr$_{0.965}$V$_{0.035}$ (blue), and
Cr$_{0.95}$V$_{0.05}$ (green) (at $15^o$ from the [110] direction)
crystals taken over a field interval of between 10 and 33 T. All data
at 4.2 K except $x=0.05$ data with field parallel to [100] which was
at 0.7 K. Insets show the data over an extended frequency range
displaying the absence of higher frequency oscillations that may be
related to the larger FS sheets of paramagnetic Cr. Drawings at the
top of figure are of a schematic representation (not to scale) in an
extended zone scheme of the magnetic breakdown orbits that occur
between intersecting hole ellipsoids formed by the remapping of the
paramagnetic Fermi surface by translation by multiples of the spin
density wave vector{\protect{\cite{fawcett,graebner}}}.  }
\end{figure}

The oscillatory magnetic susceptibility that we observe in our pulsed
field dHvA measurements is demonstrated in Fig.~\ref{fig:rawdata}a
during both the injection of current into the magnet windings and the
somewhat slower decay of the field for a nominally pure Cr crystal.
The Fourier transforms of these signals for 2 Cr crystals as well as
from a Cr$_{0.98}$V$_{0.02}$ and a Cr$_{0.965}$V$_{0.035}$ crystal is
displayed in Fig.~\ref{fig:pfft}. For our nominally pure Cr crystals,
we observe a number of narrow, well defined, dHvA frequencies between
100 and 2400 T that agree well with previous
measurements\cite{graebner,fawcett,venema,ruesink} represented in
Fig.~\ref{fig:pfft}a by the dashed lines. There are several orbits,
labeled $\chi$, $\rho$, and $\mu$ that were seen in the previous dHvA
measurements in the orientation where magnetic field is parallel to
$Q_{SDW}$ that do not appear in our data. Since our crystals were
cooled through $T_N$ in zero magnetic field, we expect our samples to
have multiple AFM domains with $Q_{SDW}$ oriented along each of the
equivalent [100] crystal axes. The absence of signals for the
frequencies associated with $Q_{SDW}$ parallel to $H$ indicates that
our experiments are somehow insensitive to these orbits. We also find
a number of smaller amplitude signals at frequencies above 2.5 kT that
we identify as either simple harmonics, or reunion orbits, of the
signals displayed in Fig.~\ref{fig:pfft}. In addition, we observe a
large number of small amplitude signals at $f<1000$ T that grow as the
temperature is reduced below 1.5 K which are not yet identified. The
large number of frequencies that appear in the dHvA spectra of Cr,
described previously as 'pseudoharmonic', are due to the
reconstruction of the FS of paramagnetic Cr by the periodicity of the
incommensurate SDW state. The remapping of the FS allows a number of
magnetic breakdown orbits which we identify in Fig.~\ref{fig:pfft}a
and in Fig.~\ref{fig:tfft} where we reproduce a schematic
demonstrating several of the cyclotron orbits of the hole surface at
N\cite{graebner,fawcett}. First, second, and third order magnetic
breakdown orbits are allowed, as well as reunion orbits. Despite the
large increase in magnetic field range afforded by the pulsed field
magnets, no signal from either the 'electron jack' or the 'hole
octahedron' portions of the FS could be identified.

Although the large field used did not result in the detection of new
oscillation frequencies related to the sheets of FS most closely
associated with the SDW state, it was sufficient to access signals
from the same FS pockets in our V-substituted crystals. This is
apparent in Fig.~\ref{fig:pfft}c and d for our $x=0.02$ and $x=0.035$
crystals where distinct frequencies of oscillation are apparent below
14 K. The low frequency spectra, $f<1 $ kT, of our $x=0.035$ crystal
is dominated by, what appear to be, noise that grows with magnetic
field. For $x=0.02$ the spectra contains a large number of peaks with
$f$ similar to that seen in pure Cr, although the peaks appear to be
broader despite the identical field range used to compute the
FFTs. The most obvious differences with pure Cr are the changes in the
relative amplitudes of the peaks and the shifting of several of the
peaks to higher frequency.  These changes are expected since, naively,
V-substitution is known to reduce the electron density and,
consequently, increase the size of the hole pocket thus reducing the
SDW wave vector. The result is that the amplitude of the dHvA
frequency associated with the unreconstructed FS, labeled $\lambda$ in
Fig.~\ref{fig:pfft}a, grows relative to the amplitude of the other
magnetic breakdown orbits and its frequency increases by $30 \pm 5$ T
per $1\%$ V substitution. There are much more dramatic changes evident
in the $x=0.035$ dHvA spectra shown in Fig.~\ref{fig:pfft}d. Here we
observe only a single frequency at 1.4 kT which is slightly larger
than that identified as the $\lambda$ orbit in frames a and b of the
figure. Thus, as we approach V concentrations where the SDW phase is
suppressed, we no longer observe the large number dHvA signals
associated with the remapping of the FS due to the formation of the
SDW state. This observation further supports the picture developed
from the earlier measurements of the dHvA in pure Cr, that of
pseudoharmonic dHvA signals resulting from the remapping of the Fermi
surface.

\begin{figure}[htb]
  \includegraphics[angle=90,width=3.0in,bb=65 290 534
  720,clip]{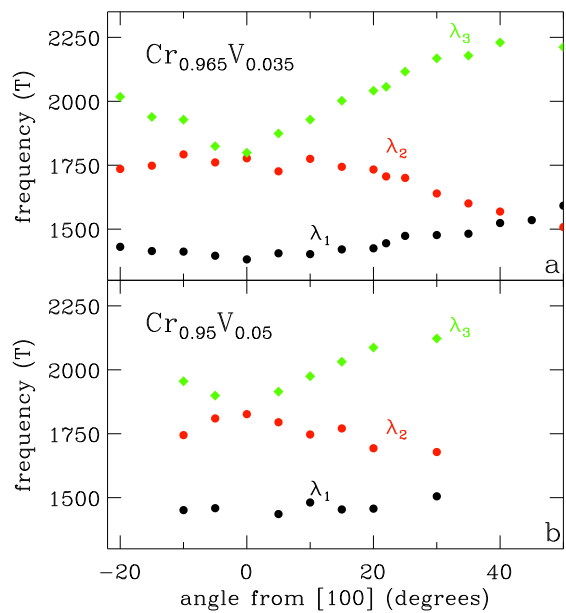}%
  \caption{\label{fig:angfig} (color online) Angle dependence of the
de Haas-van Alphen frequencies.  The variation of the three
frequencies detected in the dHvA spectra with the angle between the
crystal [100] axis and the magnetic field for rotations in the [001]
plane for a) Cr$_{0.965}$V$_{0.035}$ and b) Cr$_{0.95}$V$_{0.05}$. The
three frequency branches labeled $\lambda_1$, $\lambda_2$, and
$\lambda_3$ originate from the pocket of holes centered at the N-point
of the Brillouin zone of paramagnetic Cr. A very similar set of
frequencies is evident in the dHvA spectrum of
Mo{\protect{\cite{hoekstra}}} along with a large number of other
frequencies from the other Fermi surface sheets.}
\end{figure}

Our torque cantilever measurements of the dHvA oscillations,
demonstrated in Fig.~\ref{fig:rawdata}b, yield similar spectra as can
be seen in Fig.~\ref{fig:tfft} where we display our data for $x=0.02$,
0.035, and 0.05 crystals for two field orientations. Data taken in
this manner allows a much cleaner exploration for low frequencies
where our pulsed field data suffered from higher noise levels. The
crystals were carefully rotated in field by increments of 5 or 10$^o$
in the [001] plane for all three samples, and in the [011] plane for
the $x=0.035$ sample, to probe the variation of the Fermi surface. The
same general trends with $x$ that we observed in our pulsed field data
are also apparent in these data, the most important being the absence
of pseudoharmonic magnetic breakdown orbits for our $x=0.035$ and
$x=0.05$ samples. The variation of the 3 observed frequencies in the
dHvA spectra, labeled $\lambda_1$, $\lambda_2$, and $\lambda_3$ in
Fig.~\ref{fig:tfft}, for our $x=0.035$ and $x=0.05$ samples with
rotation of the crystal with respect to the magnetic field is
displayed in Fig.~\ref{fig:angfig}. Both the rotation symmetry of
these signals, as well as the ratio of these frequencies at high
symmetry points are nearly identical to a set of dHvA frequencies
measured in Mo\cite{hoekstra}. These frequencies are identified as
emanating from a single Fermi surface sheet, the small pocket of holes
at the N-point of the Mo Brillouin zone. Therefore, we make the same
identification for the dHvA frequencies we observe in Cr$_{1-x}$V$_x$
for $x>x_c$. We emphasize that we do not observe dHvA oscillations
that can be associated with any other pieces of the FS in any of our
samples, even in our $x=0.05$ sample which is well beyond the critical
concentration for the SDW ground state.

\section{The effective mass and scattering rates}
\begin{figure}[htb]
  \includegraphics[angle=90,width=3.0in,bb=65 115 534
  720,clip]{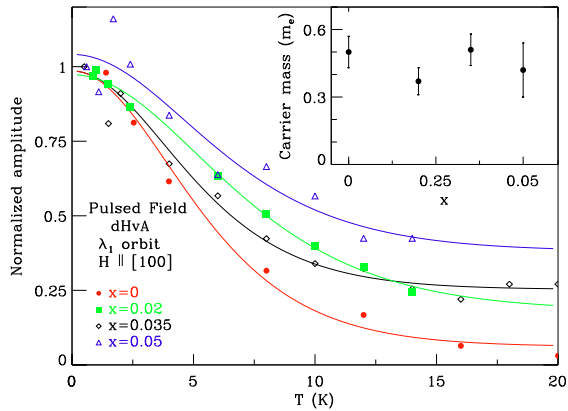}%
  \caption{\label{fig:massfig} (color online) Temperature dependence
of a de Haas-van Alphen frequency. The variation of the $\lambda$
($\lambda_1$) frequency with temperature for Cr$_{1-x}$V$_x$ as
identified in the figure with the magnetic field, $H$ oriented
parallel to the [100] crystal axis. The lines represent fits of the
Lifshitz-Kosevich formula{\protect{\cite{shoenberg}}} to the data to
determine the cyclotron mass of the charge carriers. Inset: Charge
carrier cyclotron mass as determined from the fits of the data in the
main frame plotted vs.\ vanadium concentration, $x$. No measurable
mass enhancement is observed near the critical concentration for spin
density wave ground state at $x=x_c=0.035$.}
\end{figure}

 Despite the changes to the spectra that we measure with $x$, the dHvA
signals do not have strong temperature dependencies at any $x$,
including our $x=0.035$ sample which lies closest to $x_c$. In
Fig.~\ref{fig:massfig} the amplitude of the observed dHvA frequency
identified as $\lambda$ ($\lambda_1$) in our pulsed field dHvA spectra
is plotted for the nominally pure and V substituted samples as a
function of $T$.  It is clear from this figure that there is a very
slow decay of the amplitude of the dHvA signals with $T$ for all 4
samples. The lines in the data are fits of the standard
Lifshitz-Kosevich formula\cite{shoenberg} to the data and the carrier
masses which result from these fits are plotted in the inset.  We
observe no significant mass enhancement near the QCP, or any other
trends, for the FS sheets that we are able to probe in this
experiment. This is in accord with previous determinations of the
carrier masses from dHvA in pure Cr\cite{graebner}, as well as
measurements of the linear coefficient of the specific heat in zero
field\cite{heiniger} where an increase from 1.3 to 2 mJ/mole K$^2$ was
found in going from $x=0$ to $x=0.05$, and estimates based upon
magnetic neutron scattering\cite{hayden} in V-doped samples.  Our
analysis of the amplitude reduction with temperature of the $\lambda$
orbit, in both our pulsed field and torque cantilever measurements,
reveal no significant change of carrier mass from $0.5\pm 0.1$ times
the bare electron mass of pure Cr\cite{fawcett}.

\begin{figure}[htb]
  \includegraphics[angle=90,width=3.0in,bb=55 290 543
  697,clip]{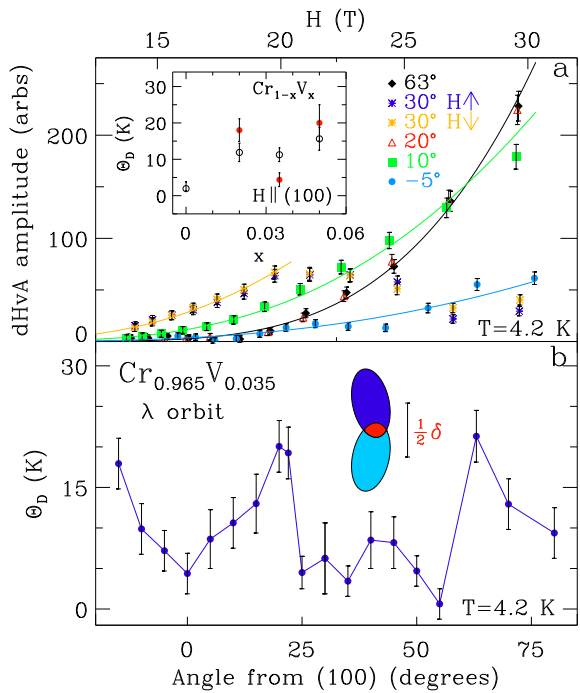}%
  \caption{\label{fig:dingt} (color online) Dingle Temperatures. a)
Magnetic field, $H$, dependence of the amplitude of the torque
cantilever de Haas- van Alphen signal for $x=0.035$ associated with
the N point ($\lambda_1$ orbit) of the bcc Brillouin zone at 4.2 K and
several angles identified in the figure. Increments in $1/H$
containing 10 full oscillations of the dHvA signal were used to
determine the amplitude plotted.  A wide variation of the Dingle
temperature, $\Theta_D$, for different $H$ orientations is apparent as
well as a strong suppression at $H > 25$ T for $H$ oriented 30$^o$
from the [100] direction. Lines are fits to the standard form for
determining $\Theta_D$, see text\cite{shoenberg}. Inset: V
concentration, $x$, dependence of $\Theta_D$ for the $\lambda$ orbit
with $H$ parallel to the [100] direction for our pulsed field
measurements, circles, and torque cantilever measurements, bullets. b)
Angle dependence of $\Theta_D$ determined from our torque cantilever
measurements. The angle refers to the orientation of $H$ with respect
to a [100] crystal axis. Inset: Schematic representation of the
N-pocket of the FS demonstrating the case where the remapping due to
translation by the spin density wave vector causes minimal overlap.}
\end{figure}

One of the features of the dHvA technique is that it allows the
determination of scattering rates, $\tau^{-1}$, for the charge
carriers residing on different parts of the Fermi surface. This is
done by measuring the amplitude of the $\lambda$ orbit as a function
of magnetic field to determine the Dingle temperature, $\Theta_D =
\hbar/2\pi k_B \tau$, where $k_B$ is Boltzmann's constant. The dHvA
amplitude, $A_p$, for the $p$'th harmonic of $f$ can be expressed as a
function of field as $A_p\propto TH^{-n}\exp{(-pZ\Theta_D)} /
\sinh{(p\alpha T/H)}$, where $n=-1/2$ for torque cantilever
measurements or $n=5/2$ for the pulsed field measurements, $Z=\alpha/
H$, and $\alpha = 2\pi^2k_Bm^*c/e\hbar$\cite{shoenberg}. In
Fig.~\ref{fig:dingt} we plot the amplitude of the $\lambda$ orbit
determined from the torque cantilever measurements for our $x=0.035$
crystal at 4.2 K as a function of field for several orientations of
the crystal. The typical form is apparent for orientations of the
field at 20 and 63$^o$ from the [100] crystal axis, whereas at some
angles significant deviations from the standard behavior is
observed. The most obvious of these deviations occurs at an angle of
30$^0$ for fields above $\sim 22$ T corresponding to the raw dHvA data
in Fig.\ref{fig:rawdata}b where we plot an arrow to indicate the field
where the amplitude deviates from the standard form. In addition, it
is clear from the changes that occur with the direction of $H$ that
there is significant variation of the carrier scattering time on this
hole ellipsoid. The lines in the figure represent fits of the above
form to the data carried out by varying the overall amplitude and
$\Theta_D$ to find the best representation and these values are
plotted in frame b of the figure.  Here we observe a wide variation of
$\Theta_D$ with the expected crystal symmetry including very large
$\Theta_D\sim 20$ K near 20 and 63$^o$ from the [100].  In contrast,
for $x=0.05$ the field dependence of the dHvA amplitudes are all
consistent with the standard form with $\Theta_D$ values scattered
within $\sim 5^o$ of 25 K.  In the inset to Fig.~\ref{fig:dingt} we
plot the $x$ dependence of $\Theta_D$ determined for fields along the
[100] direction for both measurement techniques. We observe a large
increase in $\Theta_D$ for the V substituted samples as would be
expected since the V impurities act as scattering centers for the
charge carriers.

\begin{figure}[htb]
  \includegraphics[angle=90,width=3.5in,bb=85 0 508
  697,clip]{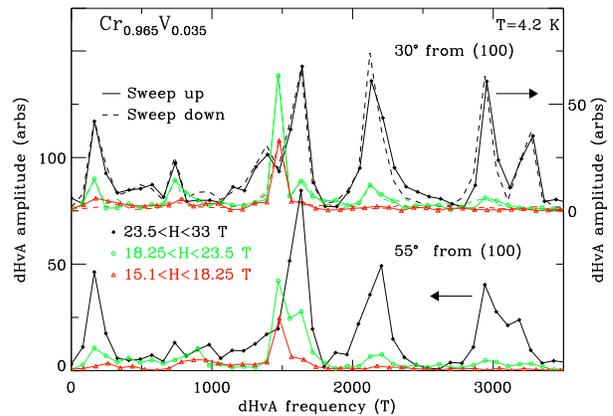}%
  \caption{\label{fig:t25} (color online) de Haas-van Alphen spectrum
  over restricted field ranges. Torque cantilever dHvA spectra for a
  Cr$_{0.965}$V$_{0.035}$ crystal for field ranges identified in the
  figure for magnetic fields oriented 30$^o$ (top) and 55$^o$ (bottom)
  from a [100]. Field ranges were chosen so that equal ranges are
  $1/H$ are displayed. }
\end{figure}

There are several possible reasons for deviations of the dHvA
amplitude from the standard form, such as we observe at an angle of
$30^o$ in Fig.~\ref{fig:dingt}a, that we have considered. This
includes magnetic breakdown, MB, and the effects of magnetic
interaction. To explore the cause for the suppression of dHvA
amplitudes with fields above 22 T, we plot in Fig.~\ref{fig:t25} the
dHvA spectra for two orientations of the magnetic field taken over the
field ranges noted in the figure. Each of these field ranges consists
of equal intervals in $1/H$ corresponding to 20 oscillations of the
$\lambda$ orbital. The reproducibility of our data is apparent in the
figure where we plot the spectra for both increasing and decreasing
field for one of the orientations. These data reveal several
interesting changes with magnetic fields near 22 T including a shift
of spectral weight toward higher frequencies from 1450 T to 1625 T.
In addition, oscillations with $f$ of 700, and 2100 T and possibly at
150 T, although the field range in the figure corresponds to only 2
full oscillations at such a small frequency, become apparent only at
higher fields.  Such changes are indicative of magnetic breakdown
orbits forming above $\sim 22$ T, which would account for our
difficulties in determining $\Theta_D$ for these orientations. These
changes are absent from our other samples on either side of the QCP
suggesting that it is a feature associated with samples with $x$ just
beyond $x_c$.

The accepted picture of Cr$_{1-x}$V$_x$ as $x$ is varied includes the
continuous modification of the FS as the charge carrier density is
varied by the substitution. The result is a decreased ordered magnetic
moment from the $0.4 \mu_B$ of pure Cr, as well as a decreased
wavevector for the SDW from 0.9513 a$^*$ for pure Cr at 4 K down to
0.926 at $x=0.025$\cite{koehler}, and inelastic fluctuations at 0.92
for $x=0.05$\cite{hayden}. The reduction of $Q_{SDW}$ with $x$
corresponds to an increased incommensurability which decreases the
overlap of the N-pocket of holes with translation by $nQ_{SDW}$ so
that the overlap becomes zero near $x=0.035$. For our sample nearest
the critical concentration, we observe magnetic breakdown in a
restricted range of orientations of the magnetic field above 22 T for
orbits that have long been associated with the N-pocket of holes. This
contrasts with the magnetic breakdown orbits observed in nominally
pure Cr, where a multitude of such frequencies is apparent even at the
lowest fields (2 T) and in a wide range of
directions\cite{graebner,fawcett,venema,ruesink}. In addition, our
dHvA spectrum for Cr$_{0.98}$V$_{0.02}$ indicates such breakdown
orbits at all fields where $\omega_c \tau > 1$ at all orientations
that we probed. Associated with the magnetic breakdown at high field
in Cr$_{0.965}$V$_{0.035}$ we find extraordinarily large $\Theta_D$
for magnetic field directions just outside of where the magnetic
breakdown orbits are observed. We speculate that we may be witnessing
a field induced SDW transition for magnetic fields above 22 T. The
dependence on magnetic field direction suggests that the field
modifies the Fermi surface in such a way as to enhance the nesting
which lies at the root of the SDW phase in pure Cr. This is similar to
the field induced SDW transition seen in the Bechgaard salt
(TMTSF)$_2$ClO$_4$\cite{chaikin,chaikin2} where magnetic fields modify
the Fermi surface so that it is more one dimensional and, thus, more
susceptible to the formation of a density wave phase. In the present
case, while the magnetic field is not likely to reduce the effective
dimensionality of the FS, we suggest that it modifies the FS in such a
manner as to increase the areas which are nearly parallel and support
nesting, but which do not change the FS volume. We note that the
Landau level filling factor for the orbits observed is of order 50 in
the field range where the MB is observed.  This filling factor is
sufficiently small so that subtle changes to the FS can be
expected. However, these changes may be significant when in close
proximity to the critical point. An example of a topological
deformation of a Fermi surface has been observed in CeB$_6$ when
subjected to strong magnetic fields\cite{harrison}. Thus, field would
tend to reverse the changes to the FS shape caused by V substitution
and, perhaps, to mimic more closely the changes to Cr that occur with
pressure\cite{feng}. This hypothesis may also explain the very large
scattering rates indicated by the high $\Theta_D$ for field
orientations just beyond where MB orbits are found. For these field
directions the modification of the FS may only be sufficient to
increase the SDW fluctuations leading to an enhanced scattering rate
for carriers. Although this hypothesis is clearly speculative, our
data are consistent with these ideas and experiments to search for a
field induced SDW phase in Cr$_{0.965}$V$_{0.035}$ are suggested.

\section{Discussion and Conclusions}
Our exploration of the Fermi surface of Cr$_{1-x}$V$_x$ for V
concentrations on both sides of the QCP at $x_c=0.035$ employing dHvA
oscillations has resulted in several important observations.  The most
conspicuous is the lack of dHvA oscillations associated with the large
sheets of the FS thought to be responsible for the SDW formation in Cr
for $x\ge x_c$. No dHvA signal from these FS sheets was observed
despite the broad range of fields ($H\le 55$ T for $x=0.035$ and $H\le
33$ T for $x=0.05$) and orientations we have probed. This is in stark
contrast to the observation of these same large FS sheets via dHvA in
Mo and W at much lower fields as well as the observation of all of the
Fermi surface sheets in Cr and Cr$_{1-x}$V$_x$ in ARPES and 2D-ACAR
measurements that are less sensitive to carrier scattering. The
absence of dHvA signals from FS sheets in Cr$_{1-x}$V$_x$ for $x \ge
x_c$ other than the pocket of holes at the N-point, could be caused by
a large scattering rate, a very large carrier effective mass, or a
combination of both. Given that the electronic contribution to the
specific heat, a Fermi surface average property, increases by less
than a factor of 2 for $x \ge x_c$\cite{heiniger} despite the factor
of 2 increase in apparent carrier density\cite{yeh,lee}, we believe it
unlikely that a large carrier mass is responsible. We are forced to
conclude that the carriers on the larger FS sheets are more
highly scattered than those associated with smaller hole FS sheet. The
lack of dHvA oscillations allows us only to calculate a lower limit
for the scattering rate of carriers residing on these sections of the
FS, $2.5\times 10^{13}$ and $1.5\times 10^{13}$ s$^{-1}$ for $x=0.035$
and 0.05 respectively. These can be compared to estimates of the
scattering rate made from transport measurements for carriers that
reside on the FS sheets that survive the SDW formation of $4\times
10^{11}$ and $5\times 10^{12}$ s$^{-1}$ for nominally pure and
$x=0.02$ single crystals respectively\cite{yeh,lee}.

It appears that although the electron jack and hole octahedron FS
sheets, much celebrated because their nesting properties are
responsible for SDW ordering, are likely restored for $x>x_c$, the
strong SDW fluctuations create a large scattering rate for the
carriers residing on these FS sheets.  This is consistent with
inelastic neutron scattering experiments that indicate significant
magnetic fluctuations even at V concentrations of
$x=0.05$\cite{hayden}. Thus, it follows that carrier scattering due to
these fluctuations, or even the opening of a pseudogap derived from
the antiferromagnetic fluctuations\cite{yeh,lee}, is responsible for
the lack of dHvA signals from regions of the Fermi surface that are
gapped in the SDW state.

Recent transport measurements on Cr$_{1-x}$V$_x$ have discovered a
dramatic change in the Hall constant as $x$ is increased through
$x_c$, or equivalently pressure was increased through the critical
pressure for the SDW AFM state\cite{yeh,lee}.  While the Hall effect
in Cr, and Cr$_{1-x}$V$_x$ for $x<x_c$, demonstrates the opening of an
energy gap over large fractions of the FS, for $x>x_c$ only the
$T$-dependent effects of fluctuation scattering remain. Surprisingly,
the changes in the resistivity are less dramatic\cite{yeh,lee}. Models
of the transport in Cr$_{1-x}$V$_x$ point out that the Hall
conductivity is rather insensitive to the changes occurring on the
flat regions of the Fermi surface, including those that are gapped as
the SDW is formed\cite{norman,bazaliy}. Instead, the changes to the
Hall constant are shown to be due to variation of the longitudinal
conductivity as the SDW phase is entered. Since the Hall constant is
dependent on the square of the longitudinal conductivity, it is much
more sensitive to the changes that occur at the SDW transition.
Nonetheless, the relatively modest changes in the resistivity at $T_N$
and at $x_c$ that are observed, despite the gapping of a large
fraction of the FS in the SDW state, can be understood if the carriers
residing on the electron jack and hole octahedron are subject to
significant fluctuation scattering in the paramagnetic state. The
contribution to the longitudinal conductivity by these sections of FS
is then small compared to that of FS sheets that are less involved in
the SDW instability. Our data suggest that the pocket of holes
centered at the N-point of the Brillouin zone experience a much
smaller scattering rate and which may consequently dominate the
carrier transport.  This allows us to speculate that the non-Fermi
liquid transport properties that are associated with magnetic QCPs in
many materials may be absent in Cr$_{1-x}$V$_x$ simply because the
scattering dependent electronic properties are dominated by FS sheets
that are relatively unaffected by the SDW transition.

Our data show that hole ellipsoid centered on the N point of the
Brillouin zone continues to dominate the dHvA spectra at $x>x_c$,
demonstrate the existence of regions of large carrier scattering rates
on this Fermi surface sheet, and hint at the possibility of a field
induced SDW phase for $x$ just beyond $x_c$. This last observation
suggests a third possibility for probing the QCP in Cr$_{1-x}$V$_x$,
application of large magnetic fields, which may be more
straightforward than varying either the V concentration or the
external pressure. Such an exploration may yield new physics and offer
insight into the importance of time reversal invariance in determining
the character of QCPs.

We are grateful to D. A. Browne and M. R. Norman for discussions. JFD
acknowledges support from the National Science Foundation through
DMR084376. A portion of this work was performed at the at the NHMFL,
which is supported by National Science Foundation Cooperative
Agreement No. DMR0654118, by the State of Florida, and by the
Department of Energy.


 \end{document}